\documentstyle[prl,aps,psfig]{revtex}

\newcommand{\be}{\begin{equation}}
\newcommand{\ee}{\end{equation}}
\newcommand{\br}{\begin{eqnarray}}
\newcommand{\er}{\end{eqnarray}}

\newcommand{\bd}{\begin{displaymath}}
\newcommand{\ed}{\end{displaymath}}

\newcommand{\bfig}{\begin{figure}}
\newcommand{\efig}{\end{figure}}

\def\3cdot{\cdot \cdot \cdot}

\def\om0{\omega _0}
\def\Om0{\Omega _0}

\def\rg{\rangle}

\def\text#1{{\rm{#1}}}

\def\->{\rightarrow}
\def\=>{\Rightarrow}
\def\-->{\longrightarrow}
\def\==>{\Longrightarrow}

\def\pr{^\prime}
\def\pr2{^{\prime\prime}}

\def\bfig{\begin{figure}}
\def\efig{\end{figure}}

\begin{document}
\draft
\title{Discrete teleportation protocol of continuum spectra field states}
\author{M. C. de Oliveira~$^1$\cite{marcos}, and G. J. Milburn~$^2$}
\address{$^1$Departamento de F\'\i sica, CCT, Universidade Federal de S\~ao Carlos,\\
 Rod. Washington Luiz Km 235, S\~ao Carlos, 13565-905, SP, Brazil.\\
$^2$~Centre for Quantum Computer Technology, Department of Physics, \\
University of Queensland, QLD 4072, Brisbane, Australia.}
\date{\today}
\maketitle
\begin{abstract}
A discrete protocol for teleportation of superpositions
 of coherent states of optical cavity fields is presented.
Displacement and parity operators are unconventionally used in Bell-like
measurement
for field states.
\end{abstract}



\section{Introduction}

When two spatially delocalised parties, A and B, share a pair of entangled
systems,
a ``quantum channel'' is established allowing information transfer (the
state of a third party
 owned by A)
 from one to the other party (A to B). Such a process is well known as
teleportation \cite{bennett}.
In fact the information about the third party state achieved by B through
the quantum channel
is maximal, originating an ambiguity of all possible outcomes, i.e., all
allowed states
spanning the Hilbert space of the object (system) whose state is to be
teleported.
 The addition of a classical channel reduces the ambiguity. With two bits
of classical information
sent from party A to party B, the latter party can decide which unitary
operation to apply in their physical system
state to recover the state
of the system to be teleported. The great concern of Einstein {\it et al.}
\cite{epr} on superluminal
 information transfer when a non-local entangled state is established is
not manifested
 by the need of a
classical channel, which validate the available information achieved
through the quantum channel.

One can describe systematically the necessary elements for efficient
teleportation:
 i) a quantum channel, i.e., a pair of nonlocally entangled systems;
ii) A Bell joint measurement of two simultaneously observable quantities
\cite{braunstein2}; and
 iii) a classical channel to transfer the information obtained in the Bell
measurement.

Subjected to these conditions many proposals have been made, following the
original proposal of Bennett et al.
\cite{bennett} considering dichotomic variables. It was first demonstrated
experimentally by
 Bouwmeester et al. \cite{bouwmeester} in a remarkable achievement, in
which a four photon
 coincident detection is necessary for a reasonable photon polarization
state teleportation.
More recently much effort has been directed to teleportation of states with
continuous spectrum,
 which basically reduces to a Wigner function representation of states and
their respective
 reconstruction. Braunstein and Kimble \cite{braunstein} proposed it
theoretically, and soon
after it was achieved experimentally \cite{furusawa}.

It is interesting to consider the extension of these ideas to more
complicated systems.
For example, is every entangled source sufficient to establish an efficient
quantum channel;
and more important,  for any system can a Bell-like measurement be made.
Given the existence
and control of an entangled pair, the search for simultaneous observables
may be a challenge.
It is important to notice, that although the original Bennett {\it et al.}
\cite{bennett} protocol was based on a complete Bell state measurement,
it was only recently achieved experimentally \cite{shih} and as far as we
know, no real complete implementation of the Bennett
protocol has been achieved up to now.
 Another interesting question, raised first by Popescu \cite{popescu},
which we discuss elsewhere \cite{mco} is:
What is the exact
 relation between Bell's inequalities violation and teleportation? Or in
other words:
 Are there states that violate Bell's inequalities but which cannot be used
for teleportation?

In this paper we pursue questions related to Bell state measurements.
We study
a specific system which allows teleportation of the state of a simple
harmonic oscillator, but through
 a protocol resembling closely the dichotomic variable protocol proposed by
Bennett
 et al. \cite{bennett}. Other attempts were made on  the discrete
formulation of teleportation
 of oscillator states \cite{vanenk,milburn,milburn2},  however these are
distinct from the original Bennett {\it et al.}
 protocol.

The element to be teleported is a quantum field state. Entanglement of
field states can be
obtained in optical networks
when a non-linear optical element is present or by linear optical elements
when one of the fields
 is prepared initially in a nonclassical state \cite{ent}.
Recently Milburn and Braunstein have addressed the problem of teleportation
when
a pair of entangled photons is generated in a parametric down conversion
scheme through
a two-mode squeezed state \cite{milburn}. Here instead, the observed
field-field correlation for conditioned phase shift
in optical cavities \cite{kimble} is used as a quantum resource.

Birefringence measurements of a single atom strongly coupled to a
high-finesse optical resonator were reported \cite{kimble}, with non-linear
phase
shifts observed in phase and probe fields for intracavity photon number much
less than one. The measured
conditional phase shifts were then proposed to be utilized for implementing
quantum
logic as a quantum-phase gate (QPG).
The possibility of using these entangled states for teleportation is here
analysed. For that we consider a model Hamiltonian to account for the
conditional phase-shift and analyze
the dynamics of the two mode states. A realisable set up based on homodyne
measurements is proposed
for the teleportation of superpositions of coherent states.

This paper is organised as follows. In Sec. II we present the generation of
a coherent entangled state
of two fields and  analyse its dynamical structure. In Sec. III we analyse
which kind of operators can be used for
 simultaneous measurement; more specifically we encounter product
combinations of displacement and parity
operators acting as the Bell operators. In Sec. IV we analyse the discrete
protocol for teleportation
 of superpositions of coherent states.
In Sec. VI we discuss the meaning of parity and displacement measurements
and propose a set up
where the Bell state measurement is considerably
simplified to homodyne detection if the target and Alice mode are first
entangled.
In Sec. VII we present a conclusion enclosing the paper.

\section{conditional dynamics}
Let us consider the dynamics generated by the following Hamiltonian
\begin{equation}
\label{a13}
H=\hbar\omega_a a^\dagger a+\hbar\omega_b b^\dagger b+\hbar \chi a^\dagger
a b^\dagger b.
\end{equation}
where $a$ and $b$ are annihilation operators for two distinct harmonic
oscillator modes,
respectively and $\chi$ is a coupling constant. Such a Hamiltonian for
optical systems describes
a four-wave mixing process, when the constant $\chi$ is then proportional
to the third order
 susceptibility \cite{milb,reid}.
It can also, for instance, describe two distinct modes interaction in Bose
condensates \cite{zoller}.
For our purpose here, it describes the effective interaction of output pump
and probe fields
of an optical cavity mediated by a two-level atom, in the dispersive limit. A strong
field-field coupling at the few photons limit, induced by non-resonant interactions between
the fields and Cs atom beams was observed experimentally \cite{kimble}.
An alternative scheme based on adiabatic transformation of a nonlinear
Hamiltonian is
described in Cochrane et al. \cite{cochrane}.
If the pumping and probe fields are prepared in coherent states,
$\left|\alpha\right\rangle_a$ and $\left|\beta\right\rangle_b$,
respectively, the evolution operator $U(t)=e^{-iHt/\hbar}$ acts over these states
as
\begin{equation}
|\psi(t)\rangle=U(t)\left|\alpha\right\rangle_{a}\left|
\beta\right\rangle_{b}=e^{-|\alpha|^2/2}\sum_{m=0}^{\infty}
\frac{\left(\alpha
e^{-i\omega_at}\right)^m}{\sqrt{m!}}\left|m\right\rangle_{a}\left|
\beta e^{-i\omega_bt}e^{-i\chi mt}\right\rangle_{b},
\end{equation}
which for $t=\pi/\chi$, turns out to be the entangled state given by
\begin{eqnarray}
\label{a4}
|\psi(\pi/\chi)\rangle &\equiv&
 \left| \alpha_+ e^{-i\pi\omega_a/\chi}\right\rangle_a \left|\beta
e^{-i\pi\omega_b/\chi} \right\rangle_{b}+
 \left| \alpha_-e^{-i\pi\omega_a/\chi} \right\rangle_a \left|-\beta
e^{-i\pi\omega_b/\chi}\right\rangle_{b}\nonumber \\
 &=&\left| \alpha e^{-i\pi\omega_a/\chi}\right\rangle_a
\left|\beta_+e^{-i\pi\omega_b/\chi}\right\rangle_{b}+
 \left|-\alpha e^{-i\pi\omega_a/\chi} \right\rangle_a
\left|\beta_-e^{-i\pi\omega_b/\chi}\right\rangle_{b}
\end{eqnarray}
where $\left|\lambda_\pm e^{-i\pi\omega_l/\chi}\right\rangle_l=
\left(\left|\lambda e^{-i\pi\omega_l/\chi}\right\rangle_l \pm
\left|-\lambda e^{-i\pi\omega_l/\chi}\right\rangle_l\right)/2$ for $l=a,\;b$
and $\lambda=\alpha,\;\beta$, respectively.
Choosing properly the modes frequency, $\omega_a$ and $\omega_b$ a set of
approximately orthogonal states can be generated as is summarized as follows
\begin{center}
\begin{tabular}{lcc}
&  &  \\
$\omega_a$ & $\omega_b$ & $|\psi(\pi/\chi)\rangle$ \\
\tableline\tableline $2\chi\;$ & $2\chi\;\; $ & $|\Phi_+\rangle$ \\
$2\chi\;$& $\chi\;\;$ & $|\Phi_-\rangle$ \\
$\chi\;$ & $2\chi\;\;$&$|\Psi_+\rangle$ \\
$\chi\;$ & $\chi\;\;$ &$|\Psi_-\rangle$\\
\tableline &  &
\end{tabular}
\end{center}
where the final states given in the third column are
\begin{eqnarray}
\label{bell}
\left|\Phi_\pm\right\rangle&=&\left| \alpha\right\rangle_a
\left|\beta_+\right\rangle_{b}\pm
\left|-\alpha \right\rangle_a \left|\beta_-\right\rangle_{b},\\
\left|\Psi_\pm\right\rangle&=&\left| \alpha\right\rangle_a
\left|\beta_-\right\rangle_{b}\pm
\left|-\alpha \right\rangle_a \left|\beta_+\right\rangle_{b}\
\end{eqnarray}

Notice that with a reformulation of the last set of states they are
written, respectively, as
\begin{eqnarray}
\label{bell}
\left|\Phi_+\right\rangle&=&\left| \beta\right\rangle_a
\left|\alpha_+\right\rangle_{b}+
\left|-\beta \right\rangle_a \left|\alpha_-\right\rangle_{b}=|\Phi'_+\rangle\\
\left|\Phi_-\right\rangle&=&\left| \beta\right\rangle_a
\left|\alpha_-\right\rangle_{b}+
\left|-\beta \right\rangle_a \left|\alpha_+\right\rangle_{b}=|\Psi'_+\rangle\\
\left|\Psi_+\right\rangle&=&\left| \beta\right\rangle_a
\left|\beta_+\right\rangle_{b}-
\left|-\beta \right\rangle_a \left|\alpha_-\right\rangle_{b}=|\Phi'_-\rangle\\
\left|\Psi_-\right\rangle&=&\left| \beta\right\rangle_a
\left|\alpha_-\right\rangle_{b}-
\left|-\beta \right\rangle_a \left|\alpha_+\right\rangle_{b}=|\Psi'_-\rangle
\end{eqnarray}
i.e., if we permute the order and rewrite the state, $|\Phi_+\rangle$ and
$|\Psi_-\rangle$ show perfect symmetry,
while $|\Phi_-\rangle$ goes to $|\Psi'_+\rangle$ and $|\Psi_+\rangle$ goes
to $|\Phi'_-\rangle$.
 This asymmetry differentiates this kind of state from
qubits written in Bell basis \cite{bennett,braunstein}.
Another point is that, actually these states are not perfectly orthogonal,
but this can be
remedied if we take large amplitude fields, $|\alpha|$, $|\beta| \gg 1$.
To shorten the notation, from now on, we will specify the states
$|\lambda_\pm\rangle_l$ as
$|\pm\rangle_l$.
\section{Parity and displacement operators measurements as resources for
teleportation}

Our first goal is to find a set of simultaneous observables for the state
$\left|\Phi_\pm\right\rangle$ and $|\Psi_\pm\rangle$.
It turns out that these operators are exactly the displacement and parity
operators.
It may be interesting
to notice that displacement and parity operators have already been combined
in the literature as
an alternative definition of the Wigner function \cite{royer,knight}.
It is straightforward to check that the parity operators
$P_a=e^{i\pi a^\dagger a}$ and $P_b=e^{i\pi b^\dagger b}$ act as
\begin{eqnarray}
\label{p}
P_a\left|\Phi_\pm\right\rangle&=&\pm|\Psi_\pm\rangle\\
P_a\left|\Psi_\pm\right\rangle&=&\pm|\Phi_\pm\rangle\\
P_b\left|\Phi_\pm\right\rangle&=&|\Phi_\mp\rangle\\
P_b\left|\Psi_\pm\right\rangle&=&-|\Psi_\mp\rangle.
\end{eqnarray}
The parity operator by itself cannot be used for our purposes, since the
above states are not
its eigenvectors.
We can however build another set of operators, observing that the
displacement operator
 $D_a(\epsilon)=e^{\epsilon a^\dagger-\epsilon^* a} $ acts on
$|\Phi_\pm\rangle$ as
\begin{eqnarray}
D_a(\epsilon)\left|\Phi_\pm\right\rangle&=&e^{iIm(\epsilon\alpha^*)}\left|
\alpha+\epsilon\right\rangle_a \left|+\right\rangle_{b}\pm
 e^{-iIm(\epsilon\alpha^*)}\left|-\alpha+\epsilon \right\rangle_a
\left|-\right\rangle_{b}\nonumber\\
&=&\cos{[Im(\epsilon\alpha^*)]}\left(\left| \alpha+\epsilon\right\rangle_a
\left|+\right\rangle_{b}\pm
 \left|-\alpha+\epsilon \right\rangle_a \left|-\right\rangle_{b}\right)\nonumber\\
&+&i\sin{[Im(\epsilon\alpha^*)]}\left(\left| \alpha+\epsilon\right\rangle_a \left|+\right\rangle_{b}\mp
 \left|-\alpha+\epsilon \right\rangle_a \left|-\right\rangle_{b}\right).
\end{eqnarray}
For very small displacements such as $|\epsilon|\ll|\alpha|$ and assuming
hereafter, without loss of generality, that $\alpha$ is a real number and
$\epsilon$ is a pure imaginary number, it follows
\begin{eqnarray}
D_a(\epsilon)\left|\Phi_\pm\right\rangle
\approx\cos{(\epsilon\alpha)}|\Phi_\pm\rangle+i\sin{(\epsilon\alpha)}|\Phi_\mp\rangle
\end{eqnarray}
and similarly
\begin{eqnarray}
D_a(\epsilon)\left|\Psi_\pm\right\rangle
\approx\cos{(\epsilon\alpha)}|\Psi_\pm\rangle+i\sin{(\epsilon\alpha)}|\Psi_\mp\rangle.
\end{eqnarray}
On the other hand the action of the displacement operator
$D_b(\lambda)=e^{\lambda b^\dagger-\lambda^* b} $, where we assume again,
without loss of generality,
$\beta$ real and $\lambda$ imaginary pure numbers, and $|\lambda|\ll|\beta|$,
\begin{eqnarray}
D_b(\lambda)\left|\Phi_\pm\right\rangle
&\approx&\cos{(\lambda\beta)}|\Phi_\pm\rangle+i\sin{(\lambda\beta)}|\Psi_\pm\rangle\\
D_b(\lambda)\left|\Psi_\pm\right\rangle
&\approx&\cos{(\lambda\beta)}|\Psi_\pm\rangle+i\sin{(\lambda\beta)}|\Phi_\pm\rangle.
\end{eqnarray}
It is then straightforward to show that
\begin{eqnarray}
P_bD_a(\epsilon)\left|\Phi_\pm\right\rangle
&=&\cos{(\epsilon\alpha)}|\Phi_\mp\rangle +i \sin{(\epsilon\alpha)}
|\Phi_\pm\rangle\\
P_bD_a(\epsilon)|\Psi_\pm\rangle&=&-\left(\cos{(\epsilon\alpha)}|\Psi_\mp\rangle
+i \sin{(\epsilon\alpha)} |\Psi_\pm\rangle\right)\\
P_aD_b(\lambda)\left|\Phi_\pm\right\rangle
&=&\pm\left(\cos{(\lambda\beta)}|\Psi_\pm\rangle +i \sin{(\lambda\beta)}|\Phi_\pm\rangle\right)\\
P_aD_b(\lambda)|\Psi_\pm\rangle&=&\pm\left(\cos{(\lambda\beta)}|\Phi_\pm\rangle
+i \sin{(\lambda\beta)} |\Psi_\pm\rangle\right)
.\end{eqnarray}
Now fixing $\epsilon \alpha=(n+1/2)\pi$, for $n=0,1,2...$ and $\lambda
\beta=(m+1/2)\pi$, for $m=0,1,2...$, we finally obtain the following
eigenvalue equations
\begin{eqnarray}
P_bD_a(\epsilon)\left|\Phi_\pm\right\rangle
&=&i (-1)^n |\Phi_\pm\rangle\\
P_bD_a(\epsilon)|\Psi_\pm\rangle&=&i(-1)^{n+1}|\Psi_\pm\rangle\\
P_aD_b(\lambda)\left|\Phi_\pm\right\rangle
&=&\pm i (-1)^m |\Phi_\pm\rangle\\
P_aD_b(\lambda)|\Psi_\pm\rangle&=&\pm i (-1)^m |\Psi_\pm\rangle.
\end{eqnarray}

As soon as they have the same eigenvector, $P_bD_a(\epsilon)$ and
$P_aD_b(\lambda)$ are simultaneous observables (with null variance), and
can be used to obtain simultaneous
information about the respective quantum state.
Once those states are entangled, it is interesting to check if this state
is a good resource for teleportation,
 in which case the state of propagating fields, or even atomic motional
states \cite{parkins} can be teleported.
 On this point the joint operators here described play a fundamental role,
as is discussed in the next section.
It is interesting to note that despite the similarity with dichotomic
variables, again
 it is not possible to match a correspondence one to one of those operators
here described
 and the Pauli spin operators. It straightforward to check that while $P_b$
in Eq. (\ref{p})
 could be correspond to $\sigma_z$, $P_a$ in Eq. (\ref{p}) would correspond
 to $\sigma_x$, and more interesting,
$P_a D_b$ corresponds then to $\sigma_x^a\sigma_x^b$ and $P_b D_a$
corresponds to $\sigma_z^a\sigma_z^b$.

\section{Discrete protocol}
For the discrete protocol we first prepare the entangled state and the
target state, $|\psi\rangle_T$ of a third party.
We consider our pair of entangled modes prepared in the $|\Phi_+\rangle$ state.
Let the target state be a superposition of coherent states
\be
\left|\psi\right\rangle_T=c_a\left|\gamma\right\rangle+c_b\left|-\gamma\right\rangle
\ee
with $|c_a|^2+|c_b|^2=1$. Then the initial state of the system will be
\be
\left|\psi\right\rangle_T\left(\left| \alpha\right\rangle_a
\left|+\right\rangle_{b}+
\left|-\alpha \right\rangle_a \left|-\right\rangle_{b}\right).
\ee
If we write $\left|\gamma\right\rangle$ and $\left|-\gamma\right\rangle$ in
terms of $\left|+\right\rangle_T$ and $\left|-\right\rangle_T$, the total
state can be written as
\begin{eqnarray}
&&\left|\Phi_+\right\rangle_{aT}\left(c_a\left|\beta\right\rangle+c_b\left|-\beta\right\rangle\right)
+\left|\Phi_-\right\rangle_{aT}\left(c_a\left|-\beta\right\rangle+c_b\left|\beta
\right\rangle\right)\nonumber\\
&+&\left|\Psi_+\right\rangle_{aT}\left(c_a\left|\beta\right\rangle-c_b\left|-\beta\right\rangle\right)
+\left|\Psi_-\right\rangle_{aT}\left(c_a\left|-\beta\right\rangle-c_b\left|\beta
\right\rangle\right)
\end{eqnarray}
Now, in the Bell measurement process on system A+T, each one of the four
terms in the above state
has $1/4$ of chance to be detected, collapsing instantaneously the state of
the party B.
Each one of these states are eigenvector of the Bell operators $P_T
D_a(\epsilon), P_a D_T(\lambda)$,
with eigenvalues $\{i,i,-i,-i\}$ and $\{i,-i,i,-i\}$, respectively. These
eigenvalues are complex as
the displacement operators are unitary. The measurement of such an operator
would need to be described by an appropriate
generalised measurement or positive operator valued measurement (POVM),
similar to the description of the complex amplitude in heterodyne measurement.
How this measurement
can be used to transfer the classical information to Bob is described below.
As usual, without such a classical information path it is impossible for
Bob to determine
its state by any other way than guessing, obtaining the classical limit for
teleportation of $1/4$.
Let us consider that the first measurement made is described by the
operator $P_T D_a(\epsilon)$.
If the outcome is $+i$ the state is $\left|\Phi_\pm\right\rangle$;
if it is $-i$ then the state is $\left|\Psi_\pm\right\rangle$. Then the
second measurement described by $P_a D_T(\lambda)$ is made.
If the first measurement made was $+i$ then the second measurement will
give $+i$ for the state $\left|\Phi_+\right\rangle$ and $-i$ for the state
$\left|\Phi_-\right\rangle$.
Now, if the first measurement made was $-i$ then the second measurement
will give $+i$ for the
 state $\left|\Psi_+\right\rangle$ and $-i$ for the state
$\left|\Psi_-\right\rangle$.
In possession of this information, Bob can effect the necessary inverse
transformations once he has one of the following states,
\begin{eqnarray}\label{unitary}
&\left|\psi\right\rangle_b&\\
P_b&\left|\psi\right\rangle_b&\\
iD_b(\mu)&\left|\psi\right\rangle_b&\\
iP_bD_b(\mu)&\left|\psi\right\rangle_b&
\end{eqnarray}
for $\mu \beta=\pi/2$, completing the teleportation protocol. Notice that,
despite that the parties states are essentially coherent states and the
entanglement in a continuous basis,
 the teleportation scheme is analogous to the original dichotomic variables
teleportation
 protocol of Bennett {\it et al}\cite{bennett} even though the joint
operator that plays the role
of the Bell operator is a unitary non-Hermitean operator.
$|\psi\rangle_T$ can vary from a simple coherent state to a coherent
superposition, or even in the case
in which the superpositions are the even and odd coherent states, for low
intensity, $|\psi\rangle_T$ corresponds to
zero and one Fock states, respectively.

\section{Measurement of parity and displacement}

Although the formal scheme presented in the last section allows the
complete Bell state
measurements, it was not explained how it could be
realised. In fact the measurement process is divided in two stages as there
are two operators involved, the parity and displacement operators.
The parity is more to be understood as an operation over the joint field
state simultaneously to the displacement.
For field parity measurements we have to resort to the methods well
explored in microwave cavities \cite{brune}.
An atom is prepared in a superposition $(|e\rg+|g\rg)/\sqrt{2}$ and let to
interact dispersively with the field.
After the interaction the atomic state is rotated again by a $\pi/2$ pulse.
Due to the dispersive interaction, only the atom in the state $|e\rg$
causes a parity flip in the field state and finally the parity of the field
can be deduced by the detected atomic state.
However, for the operation considered here, the atom has to be prepared in
the $|e\rg$ state
and we do not read their final state.
In this way excited atoms cause a  $\pi$ shift mode state (e.g.
$|\alpha\rg\rightarrow|-\alpha\rg$).
The displacement measurement is directly given by quadrature
($X=a+a^\dagger$) measurement through homodyne detection.
As the parameter $\varepsilon$ is known to be very small, the
displacement operator, e.g. for the mode A, is given by
\be
D_a(\varepsilon)=e^{i|\varepsilon|\hat X}\approx 1+i\varepsilon
X;\;[D_a(\varepsilon), X]=0.
\ee
Knowing $|\varepsilon|=(n+1/2)\pi/|\alpha|$, the measurement of $\hat X$
gives the displacement.
 As expected the degree of control in this kind of measurement has to be
very high.

The measurement stage can be simplified dramatically if  another element is
introduced
 in the protocol,
if we actually entangle the target field with the Alice mode by the same
scheme used to generate the entangled
pair A-B. Allowing the interaction time to be again $t=\pi/\chi$ and
setting the mode frequencies as explained in the Sec. III,
 it is straight to obtain the following entangled state for the joint
system A-B-T
\br
&&\frac 1 2
\left\{|\gamma\rg|\alpha\rg(C_a|\beta\rg+C_b|-\beta\rg)+|\gamma\rg|-\alpha\rg(C_
a|\beta\rg-C_b|-\beta\rg)\right.\nonumber\\
&&\left.+|-\gamma\rg|\alpha\rg(C_a|-\beta\rg+C_b|\beta\rg)+|-\gamma\rg|-\alpha\rg(-C_a|-\beta\rg+C_b|\beta\rg)\right\}
\er
which states can be distinguished by simple homodyne detection of modes A
and T as schematically described in Fig. 1.
We should remark that we choose to deal with states with a real complex
amplitude.
In consequence the homodyne detection of each mode
gives the phase  quadrature $X$, which is distinguished in each case by
positive or negative signals. This information is
communicated  by a standard classical channel to Bob, who possesses one of
the states of Eq. (\ref{unitary})
and has to apply the respective inverse unitary operation
to obtain the original target state, completing the protocol.
\section{Conclusion}
We have discussed a teleportation protocol for harmonic oscillator states
based on
a different entanglement resource to that usually considered. The standard
teleportation resource for
an oscillator is a two mode squeezed state\cite{braunstein,milburn}.
Here we consider a teleportation resource based on
entangled coherent states generated, for example, by a Kerr nonlinearity, or by the effective coupling
of the probe and pumping field strongly coupled to an Cs atom, as observed experimentally in \cite{kimble}.
The protocol can be made
equivalent to the two qubit scheme originally proposed by Bennett et
al.\cite{bennett} and we have explicitly
identified the equivalent Bell basis measurements. In our case these would
correspond to
generalised measurement but can be realised to a good approximation as
measurements of parity and quadrature phase amplitude. It is hoped that
this alternative
teleportation protocol will prove useful in elucidating the more general
issue of entanglement between
two systems with infinite dimensional Hilbert spaces.

\acknowledgments{MCO would like to acknowledge A. S. Parkins for fruitful
discussions and
the Funda\c{c}\~ao de Amparo \`a Pesquisa do Estado
de S\~ao Paulo (Brazil) for financial support.
}


\newpage
\centerline{\bf Figure caption}

Fig. 1- Schematic of the cavity QED experimental apparatus for
teleportation of field states. A Cs Beam entangle the vertical (Bob) and horizontal (Alice) pulses.
A polarizing beam-splitter (pbs) splits the pulse in two components. The horizontal component is
entangled with the target pulse. Results of homodyne measurements made on Alice and target are sent
to Bob by classical channels.
\end{document}